\documentclass[12pt,preprint]{aastex}
\usepackage{t1enc}
\usepackage{graphicx}
\usepackage{epstopdf}
\usepackage{natbib}
\usepackage{lscape}
\usepackage{lmodern}
\newcommand{\kms}{km~s$^{-1}$}

\newcommand{\lsun}{$L_{\odot}$}

\newcommand{\cmt}{cm$^{-3}$}

\newcommand{\jpb}   {$\rm Jy~beam^{-1}$}

\newcommand{\gap}%
{\raisebox{-0.5ex}{$\stackrel{\scriptstyle >}{\scriptstyle \sim}$}}
\newcommand{\iras} {IRAS~18162$-$2048}
\newcommand{\Jdv} {J1819$-$2036}

\begin{document}

%\slugcomment{To be Submitted to the \apj. Printed \today}
\shorttitle{Proper motions in the HH 80N/81/80N radio-jet}
\shortauthors{Masqu\'e et al.}

\title{Proper motions of the outer knots of the HH 80/81/80N radio-jet}

\author{
Josep M. Masqu\'e\altaffilmark{1,2},
Luis F. Rodr\'iguez\altaffilmark{1},
Anabella Araudo\altaffilmark{3},
Robert Estalella\altaffilmark{4},
Carlos Carrasco-Gonz\'alez\altaffilmark{1},
Guillem Anglada\altaffilmark{5},
Josep M. Girart\altaffilmark{6}, and
Mayra Osorio\altaffilmark{5}
} 

\altaffiltext{1}{Instituto de Radioastronom\'ia y Astrof\'isica, Universidad Nacional Aut\'onoma de M\'exico, Morelia 58089, M\'exico}
\altaffiltext{2}{Departamento de Astronom\'ia, Universidad de Guanajuato, Apdo. Postal 144, 36000 Guanajuato, M\'exico}
\altaffiltext{3}{University of Oxford, Astrophysics, Keble Road, Oxford OX1 3RH, UK}
\altaffiltext{4}{Departament d'Astronomia i Meteorologia and Institut de Ci\`encies del Cosmos (IEEC-UB), Universitat de Barcelona, Mart\'i i Franqu\`es 1, E-08028 Barcelona, Catalunya, Spain}
\altaffiltext{5}{Instituto de Astrof\'isica de Andaluc\'ia (CSIC), Apartado 3004, E-18080 Granada, Spain}
\altaffiltext{6}{Institut de Ci\`encies de l'Espai (CSIC-IEEC), Campus UAB, Carrer de Can Magrans, S/N, 08193 Cerdanyola del Vall\`es, Catalunya, Spain}

\begin{abstract}

The radio-knots of the HH 80/81/80N jet extends from the HH 80 object to the recently discovered Source 34 and has a total projected jet size of 10.3~pc, constituting the largest collimated radio-jet system known so far. It is powered by the bright infrared source \iras\ associated with a massive young stellar object. We report 6~cm JVLA observations that, compared with previous 6~cm VLA observations carried out in 1989, allow us to derive proper motions of the HH 80, HH 81 and HH 80N radio knots located about 2.5~pc away in projection from the powering source. For the first time, we measure proper motions of the optically obscured HH 80N object providing evidence that this knot, along with HH 81 and HH 80 are associated with the same radio-jet. We also confirm the presence of Source 34, located further north of HH 80N, previously proposed to belong to the jet. 
We derived that the tangential velocity of HH 80N is 260~\kms\ and has a direction in agreement with the expected direction of a ballistic precessing jet.  The HH 80 and HH 81 objects have tangential velocities of 350 and 220~\kms, respectively, but their directions are somewhat deviated from the expected jet path. The velocities of the HH objects studied in this work are significantly lower than those derived for the radio knots of the jet close to the powering source (600-1400~\kms) suggesting that the jet is slowing down due to a strong  interaction with the ambient medium. 
As a result, since HH 80 and HH 81 are located near the edge of the cloud, the inhomogeneous and low density medium may contribute to skew the direction of their determined proper motions.  The HH 80 and HH 80N emission at 6 cm is, at least in part, probably synchrotron radiation produced by relativistic electrons in a magnetic field of 1~mG. If these electrons are accelerated in a reverse adiabatic shock, we estimate a jet total density of $\lesssim1000$~cm$^{-3}$. All these features are consistent with a jet emanating from a high mass protostar and make evident its capability of accelerating particles up to relativistic velocities.

\end{abstract}

\section{Introduction}

The Herbig-Haro (HH) objects 80 and 81 \citep{reipurth1988} are among the brightest HH
objects known. They lie at the southern edge of the dark
molecular cloud L291 in Sagittarius, located at 1.7 kpc of distance \citep{rodriguez1980}, and 
are powered  by the bright infrared source \iras. Because of its high luminosity ($1.7 \times 10^4$~\lsun), this source is believed to be associated with a massive protostar, or cluster of protostars \citep{aspin1992}. The separation
between HH 80/81, and \iras\ is 2.3~pc, the latter source having associated centimetric emission that appears as a thermal radio-jet pointing to HH 80 and 81 \citep{marti1993}. The central source
is also associated with several molecular lines and compact mm continuum emission
\citep{gomez2003, fernandezlopez2011a, fernandezlopez2013} and has
associated a disk with a radius of a few hundreds of AU
\citep{carrascogonzalez2012,fernandezlopez2011b}. About 3~pc north of
the central source there is HH 80N, a radio source that probably
constitutes the northern counterpart of HH~80 \citep{marti1993}. If HH 80N is actually an HH object, its non-detection at optical wavelengths could be explained due to this object been deeply embedded in the molecular cloud. Further evidence of its HH nature is provided by the photochemical effects detected in a condensation of molecular gas and dust located ahead of HH 80N \citep{girart1994, girart1998, masque2009}. 
More recently, \citet{masque2012b} reported a new radio source (Source 34) further north of HH 80N,
closely coinciding with the direction of the jet. The fit of a precession jet model suggests that
this source likely belongs to the HH complex (hereafter, the HH 80/81/80N jet and, hence, that the jet extends beyond HH 80N. If confirmed, this would imply that the jet has a total projected length of 10.3 pc.
 
 The characteristics of HH 80/81/80N make this radio-jet a scaled up version of the jets emanating from low-mass protostars. Indeed, proper motion measurements yield projected velocities in the plane of sky between 600 and 1400~\kms\ for the inner knots located a few arcsec away from the central source, and about 500~\kms\ for newly ejected condensations that traveled less than 1~arcsec in projection \citep{marti1995,marti1998}. Also, HH 80 and HH 81 proper motions were measured by \citet{heathcote1998} using high spatial resolution optical observations that resolved the HH objects into 
several bright condensations. The measured proper motions presented a large dispersion of velocities (from 0 to 900~\kms). In particular,
proper motions of about 300~\kms\ were found for
the brightest condensations in HH 80 and HH 81.

The interaction of the jets with the ambient medium produces shocks
 that can accelerate 
electrons up to relativistic energies and subsequently radiate synchrotron 
emission \citep{araudo2007}. This has been confirmed by the detection of 
linearly polarized radio emission at distances of $\sim0.5$~pc from the 
central source of the HH 80/81/80N jet
\citep{carrascogonzalez2010}. At larger distances, the sources
HH 80, HH 81 and HH 80N are also likely synchrotron emitters because of 
their negative spectral indices ($S_{\nu}\propto \nu^{-0.3}$, being $S_{\nu}$ 
the flux at the frequency $\nu$), similarly to the radio-knots of jets 
found in Serpens 
\citep{rodriguez1989b, rodriguez2005, curiel1993},
Cep A \citep{garay1996} and W3(H$_2$O) \citep{wilner1999}.
 The presence of hot plasma and relativistic electrons downstream the 
shock makes HH jets potential (thermal) X- and (non-thermal) 
$\gamma$-ray emitters
\citep{araudo2012,boschramon2010,raga2002}. In this sense, \citet{pravdo2004}
detected thermal X-ray emission in HH 80 and HH 81 that was claimed to 
arise when the jet slams into the ambient material.

In this paper we report proper motion measurements, made at radio wavelenghts for the first time, of the radio-knots HH~80, HH~81 and HH~80N, in order to confirm, first, that the three knots belong to the same jet and, second, that these HH objects
have significantly slower proper motions than the radio knots 
located a few arcsec from the central source, indicating strong  jet interactions with the ambient cloud. These results provide an important complement to previous proper motion studies of the HH 80/81/80N jet near the center \citep{marti1995,marti1998}, as we investigate the jet behavior far away from the central source.
Also, the results reported here allow us to estimate the magnitude of the magnetic field in equipartition with non-thermal particles ($B_{\rm eq}$) and the jet density ($n_{\rm j}$) at the position of the  HH objects. We also report observations of the region northward of HH 80N confirming the presence of Source 34 detected by \citet{masque2012b}.

\section{Observations and Results}

The new observations were carried out on 2013 September 6 using the VLA interferometer in the hybrid CnB configuration. We observed in the C band (6~cm wavelength, from 4.5 to 6.5 GHz) in continuum imaging mode with full polarization. 
We used the 8-bit
sampler and set 3 sec as the integration time. The flux and gain
calibrators were 3C286 and J1820$-$2528, respectively. The bootstrapped
6~cm flux density of J1820$-$2528 was $0.824 \pm 0.002$~Jy at the frequency of 5.36~GHz (spectral index of $0.19 \pm 0.02$). We observed 3 fields,
covering all the jet length (i.e. from Source 34 to HH 80, see
Figure~2 of \citealt{masque2012b}). The northernmost field was
centered on the highly luminous radio-source \Jdv\ (see
\citealt{marti1993} and \citealt{masque2012b}), while the two southern
fields were chosen with the aim of reproducing the pointings performed in
\citet{marti1993} (i.e. centered $\sim2'$ north and south, respectively, from the central source). The on- source time was three minutes per pointing
with the exception of the \Jdv\ field, where only two minutes were
employed.  

The data were reduced using the Common Astronomy Software Applications
(CASA) and the map was constructed with the CLEAN task with natural
weighting. The resulting beamsize was $3\farcs9 \times 3\farcs2$
(P.A. $= 44.5\arcdeg$) and the achieved $rms$ noise of the maps was 53~$\mu$\jpb\ for the two southern fields. 
For the \Jdv\ field, the strong side-lobes of \Jdv\ limited the dynamic range and increased the $rms$ noise level to $\sim500~\mu$\jpb. To solve this, we obtained a model of \Jdv\ from the
calibrated data-set and used it to self-calibrate the visibilities of
the corresponding field. The \Jdv\ source presents strong time variability (on a scale of few hours) and a spectral index of $\sim-0.5$. The integration time of the observations was shorter than the variability timescale of the source and
the self-calibration was hardly affected by this. However, the \Jdv\ flux varied importantly in the 2~GHz bandwidth. Thus, for the self-calibration process, in the CLEAN task we employed three terms of the Taylor polynomial used to model the frequency structure across the bandwidth (nterms=3). The $rms$ noise of the \Jdv\ field clearly improved reaching a value of $\sim50~\mu$\jpb. The three fields were combined into a single mosaic using the \emph{linearmosaic} function.

We also calibrated and imaged the 6~cm archival data of \citet{marti1993},
observed in September 1989 with the VLA in the C configuration. The
details of the observations are given in that paper. We used CASA in a
similar fashion than the observations presented above, but choosing
uniform weighting for the final map. This gives a synthesized beam of
$5\rlap.{''}8 \times 1\rlap.{''}9$ (P.A. = -14$^\circ$), more similar than the angular resolution of the 2013 observations, and an $rms$ noise of $40~\mu$\jpb. This noise level is smaller than that of  
the \citet{marti1993} 6~cm map and more appropriate for astrometric studies.

\begin{figure}[thbp]
\begin{center}
\vspace{-1.0cm} 
\resizebox{1.8\textwidth}{!}{\includegraphics{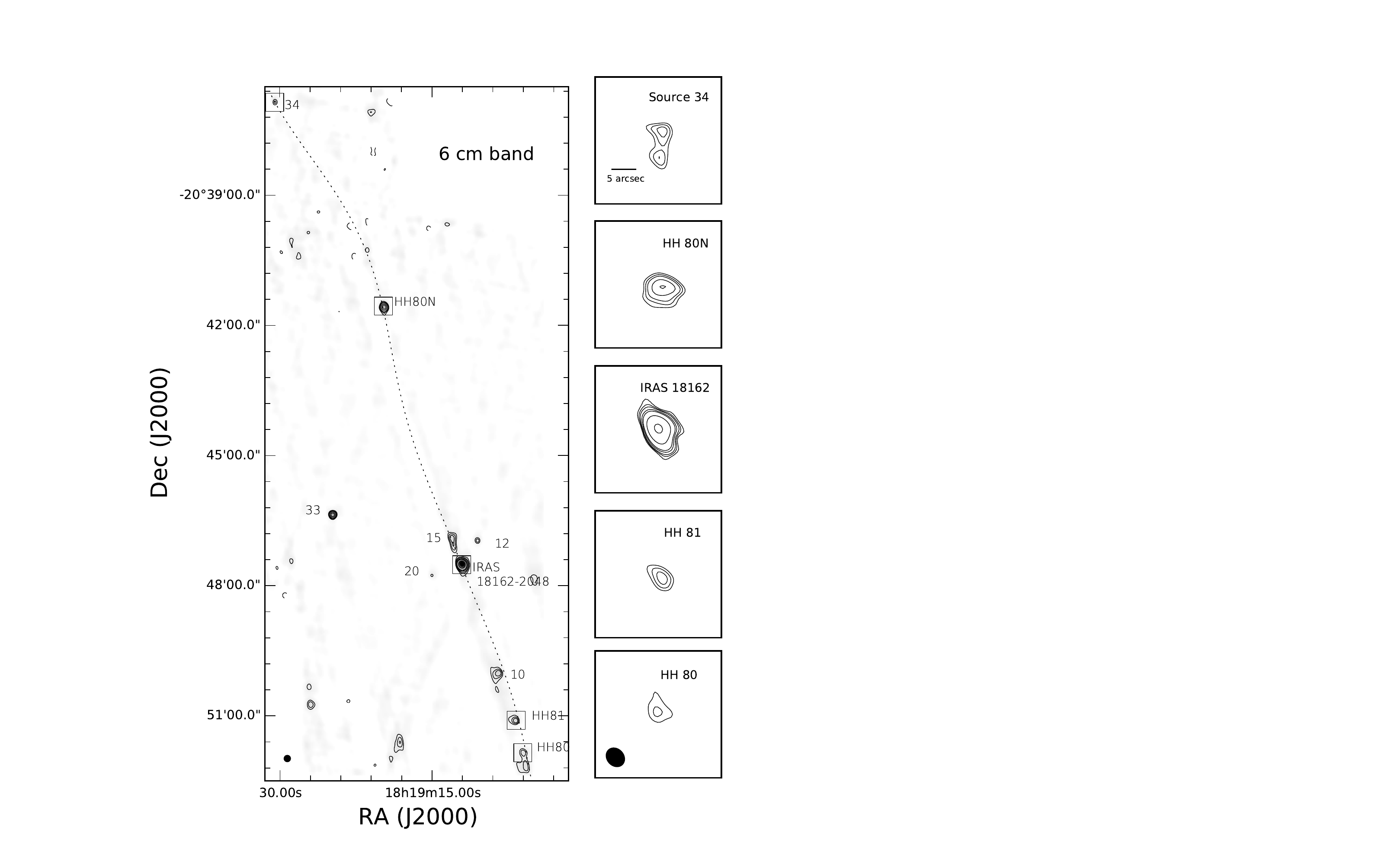}}
\vspace{-0.7cm} 
\caption{
\scriptsize \emph{Left panel:} Map at 6 cm of the HH 80/81/80N jet observed in 2013 with the JVLA in the CnB configuration convolved to a beamsize of 8\arcsec. Contours are -4, 4, 5, 6, 8, 10, 20, 50 and 100 times the $rms$ noise level of the map ($19~\mu$\jpb).
The radio knots are labeled following the nomenclature of \citet{marti1993} and \citet{masque2012b}. 
The squares show the field of view displayed in the right panels (20~arcsec). The dotted line represents the 
best fit model of a precessing jet (model B of \citet{masque2012b}). \emph{Right panels:} Zoom of the 6 cm map obtained with natural weighting from the 2013 observations, which gives a synthesized beamsize of $3.9\arcsec \times 3.2\arcsec$ (PA = 44.5$^\circ$), of (from top to bottom) Source 34, HH 80N, \iras, HH 81 and HH 80 \label{jet}. Contours are -4, 4, 5, 6, 8, 10, 20, 50, and 100 times the $rms$ noise of the map ($55~\mu$\jpb). For Source 34 and HH 80 contours -3 and 3 have been added. The scale of the maps is shown in the top panel.}
\end{center}
\end{figure}

Figure~\ref{jet} shows the 6~cm emission of the
HH~80/81/80N jet observed in 2013. Most of the sources found in \citet{marti1993} are detected, including the ones belonging to the jet. 
The 6~cm map also shows emission at the position of Source 34 , with a signal to noise ratio of 5 previously reported by \citet{masque2012b} and proposed to belong to the HH 80/81/80N jet. This independent detection confirms the presence of this source. As seen in the top right panel of
Figure~\ref{jet}, Source 34 appears splitted into two sources aligned roughly in
the north-south direction.

In Table 1 we present the results of Gaussian fits to the main radio-knots of the HH 80/81/80N jet (except for Source 34) from the 6 cm map obtained in 2013 convolved to the final beam size of $6\rlap.{''}0 \times 4\rlap.{''}0$
(PA = -14$^\circ$), which is more appropriate for a comparison with the \citet{marti1993} 6~cm map. Comparing the fluxes of this table with those of \citet{marti1993} we assess a flux drop of $\sim40$\% for HH 81, clearly above the flux calibration uncertainty of 10\%, even taking into account the different bandwidth of the data-sets. The HH 80 object has an integrated flux roughly constant in the time baseline of 24~yr. However, as seen in the lower right panel of Fig.~\ref{jet}, the 2013 map displays HH 80 slightly more extended to the southwest that is balanced by a decrease in the intensity peak with respect to the 1989 map.  Both morphological and flux variations are common in the HH objects and are attributed to changes in the internal structure of the shock \citep[e.g. HH 1--2][]{rodriguez2000}. On the other hand, the central source does not show significant variability as is typical for the central sources of thermal radiojets \citep[e.g.][]{anglada1996a}.

The proper motions of the radio-knots of the HH 80/81/80N jet were
derived from the 6 cm maps of the 1989 and 2013
data convolved to the same final beam ($6\rlap.{''}0 \times 4\rlap.{''}0$; P.A. = -14$^\circ$). Then, the fields corresponding to the northern and southern pointings of \citet{marti1993} (the \Jdv\ field has no 1989 observations) were aligned separately using a selected set of reference sources in the field detected in both epochs, not belonging to the jet (i.e. assumed to be stationary), and detected with a S/N ratio greater than 9. This is a better choice than adopting the position of the central source as reference, which has periodic ejections of new condensations that can modify slightly its morphology affecting the fitting position of the centroid of this source. For the northern field,
sources 12, 20 and 33 were used as reference while for the southern field, the same sources plus the sources 2 and 16 were used (these sources follow the nomenclature of \citealt{marti1993} and \citealt{masque2012b}). These latter sources are located outside the field of view of the left panel of Fig.~\ref{jet} and some arcminutes away from the pointing center. In order to prevent a bad fit position due to source smearing, we corrected the maps for the response of primary beam. Sources 2 and 16, despite falling close to the noisy edge of the primary beam, are bright enough to be detected with a good signal to noise ratio. The position of the sources was determined by Gaussians fits.
Then, averaging the derived positions for the reference sources we got 
a reference position for each epoch and
pointing. Using this technique we found shifts between the maps of the two epochs ($\Delta$x, $\Delta$y) of ($-0\rlap.{''}44$,$-0\rlap.{''}24$) and ($-0\rlap.{''}36$,$-0\rlap.{''}07$) for the northern and southern fields, respectively. The $rms$ of the difference in position ($\delta$x, $\delta$y) of the
reference sources between the 1989 and 2013 epochs, also known as the
alignment error, were ($0\rlap.{''}20$, $0\rlap.{''}14$) and
($0\rlap.{''}18$, $0\rlap.{''}27$) for the northern and southern
fields, respectively.

\begin{figure}[thbp]
\begin{center}
%\vspace{-1.0cm} 
\resizebox{1.0\textwidth}{!}{\includegraphics{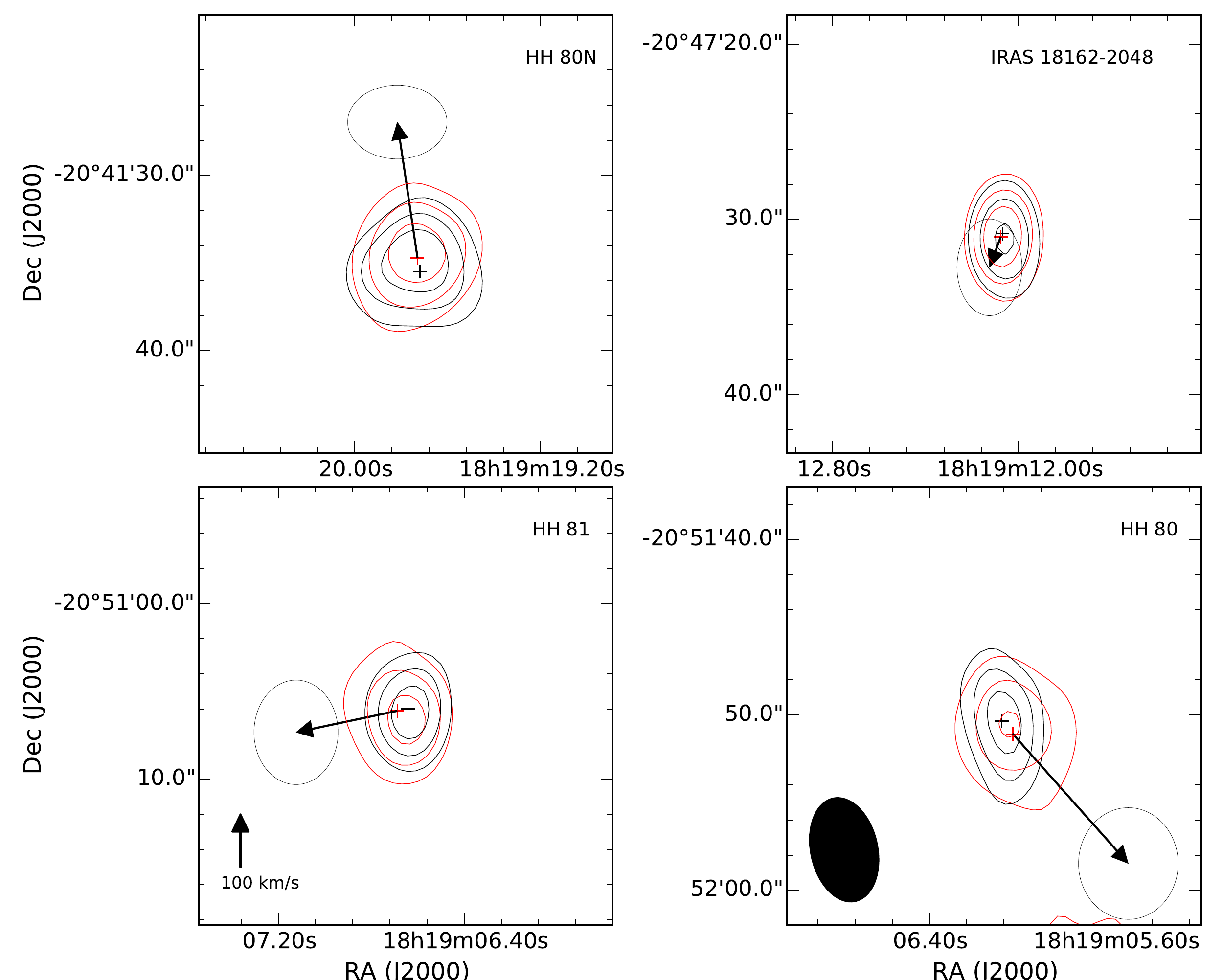}}
%\vspace{-0.7cm} 
\caption{6 cm emission corresponding to the 1989 observations (black contours) and 2013 observations (red contours) of HH 80N (top left panel), \iras\ (top right panel), HH 81 (bottom left panel) and HH 80 (bottom right panel). The maps have been convolved to the same final beam ($6.0\arcsec \times 4.0\arcsec$; PA = -14$^\circ$) that is shown in the bottom left corner of the bottom right panel. The contours are 50, 70 and 90\%\ times the representative peak value of the object (HH 80N: 0.8~m\jpb; \iras: 3~m\jpb; HH 81: 1.0 (1989) and 0.5 (2013)~m\jpb; HH 80: 0.50 (1989) and 0.39 (2013)~m\jpb). The arrows represent the magnitude and direction of the derived proper motions and the ellipses represent the uncertainties.
\label{proper_mov}}
\end{center}
\end{figure}

Once the reference positions of the 1989 and 2013 epochs were determined, the same Gaussian fitting of above was employed to measure
the relative positions of the radio-knots of the jet from the reference position of their respective field. Finally, we compared the relative
positions of the radio-knots between the two epochs obtaining their
displacement in 24 yr. To derive the total error of the displacement of the sources, the uncertainties in the fit of the position
determination of the two epochs, typically between $0\rlap.{''}1$ and
$0\rlap.{''}2$, were added
quadratically to the alignment error. To obtain the total uncertainty in the source
displacement along the PA of the motion, as well as
the uncertainty of the PA itself, error propagation was applied. The two aligned images showing the 
observed angular displacement between the two epochs of observation (separated by 24~yr) of the most important objects of the jet are
presented in Fig.~\ref{proper_mov}. The derived motions are represented by arrows, with the ellipses corresponding to the error of the motion, whose values are presented in the Table~\ref{propermotions}. From the measured displacement, we derive proper motions values ranging between 7.9 and 43.8 mas~yr$^{-1}$.

%{\bf There is the caveat that the apparent proper motions could be caused by morphological changes instead of a bulk motion of the object. 
%However, the emission of these HH objects is expected to come from the reverse shock (see discussion). If the reverse shock is stationary, changes of its emission distribution are not expected. } 

Our tangential velocities derived for the radio sources HH 80 ($351 \pm 104$~\kms), HH 81 ($223 \pm 85$~\kms)
and HH 80N ($263 \pm 71$~\kms) have similar values to those of the brightest knots of the southern HH
objects, HH 80A ($334 \pm 23$~\kms) and HH 81A ($370
\pm 17$~\kms) measured by \citet{heathcote1998} from optical observations. As seen in Fig.~\ref{proper_mov}, the direction of the motion of HH 80N is in good agreement with the expected direction for a ballistic precessing jet. The HH 80 proper motion is somewhat deviated to the west with respect to the jet path but still traveling away from the central source. On the other hand, the direction of the HH 81 proper motion is clearly not tangent to the jet
path. The proper motions derived by \citet{heathcote1998} for individual condensations in HH 80 and 81 also show considerable dispersion in velocity and direction. Finally, within the uncertainties, we find no significant displacement for the central source, as expected.

\section{Discussion}

\subsection{The HH nature of HH 80N}

Up to date, the likely HH nature of HH 80N and its association with the jet emanating 
from \iras\ was based on its  position aligned with the jet knots, 
with a very low probability for HH 80N being a background
source. In this work, we report for the first time proper
motions of HH 80N. The proper motions point in a direction consistent with a ballistic
motion of a jet element (see Fig.~\ref{proper_mov}). 
This provides evidence that HH 80N, HH 80 and HH 81 
belong to the same jet, constituting the HH 80/81/80N jet system.
Furthermore, the similarity of the HH 80, 81 and 80N spectral indices \citep[$\sim-0.3$,][]{marti1993,masque2012b}, their
comparable tangential velocities (see Table~1) and
approximate symmetric positions with respect to the central source
suggest that their emission is generated under the same mechanism. Moreover, they were likely produced in the same major ejection event of the driving 
source.

\subsection{Synchrotron emission from the HH objects}

The emission of the inner knots studied by \citet{marti1995, marti1998}, corresponding to recent ejections, is produced by internal shocks of the jet. Thus, their measured velocity is representative of the jet velocity. On the other hand, the emission of HH 80 and 81 corresponding to older ejections arises from the interaction of the jet with the ambient cloud, owing to their known HH nature. As we mentioned in previous section, HH 80N has likely the same nature as HH 80 and 81 and, hence, its emission is expected to be produced also by the jet interaction with the medium. The qualitative difference between the measured 
tangential velocities of the outer radio-knots HH~80, HH~81
and HH~80N ($\sim223$--351~\kms) and the inner radio knots ($\sim600$--1400~\kms, \citealt{marti1995}) implies that the ambient material, at least in some parts of the cloud, contributes significantly in slowing down the jet material. This is supported by the high bolometric luminosity of the outer HH objects that suggests strong  
interactions of a high velocity flow with stationary obstacles in the cloud.
Certainly, the fading luminosity of HH 81 and the change in morphology of HH 80, as it escapes from the cloud and possibly expands into a low pressure medium, makes evident the important role of the ambient cloud in the emission of these HH objects.

\begin{figure}[thbp]
\begin{center}
%\vspace{-1.0cm} 
\resizebox{1.0\textwidth}{!}{\includegraphics{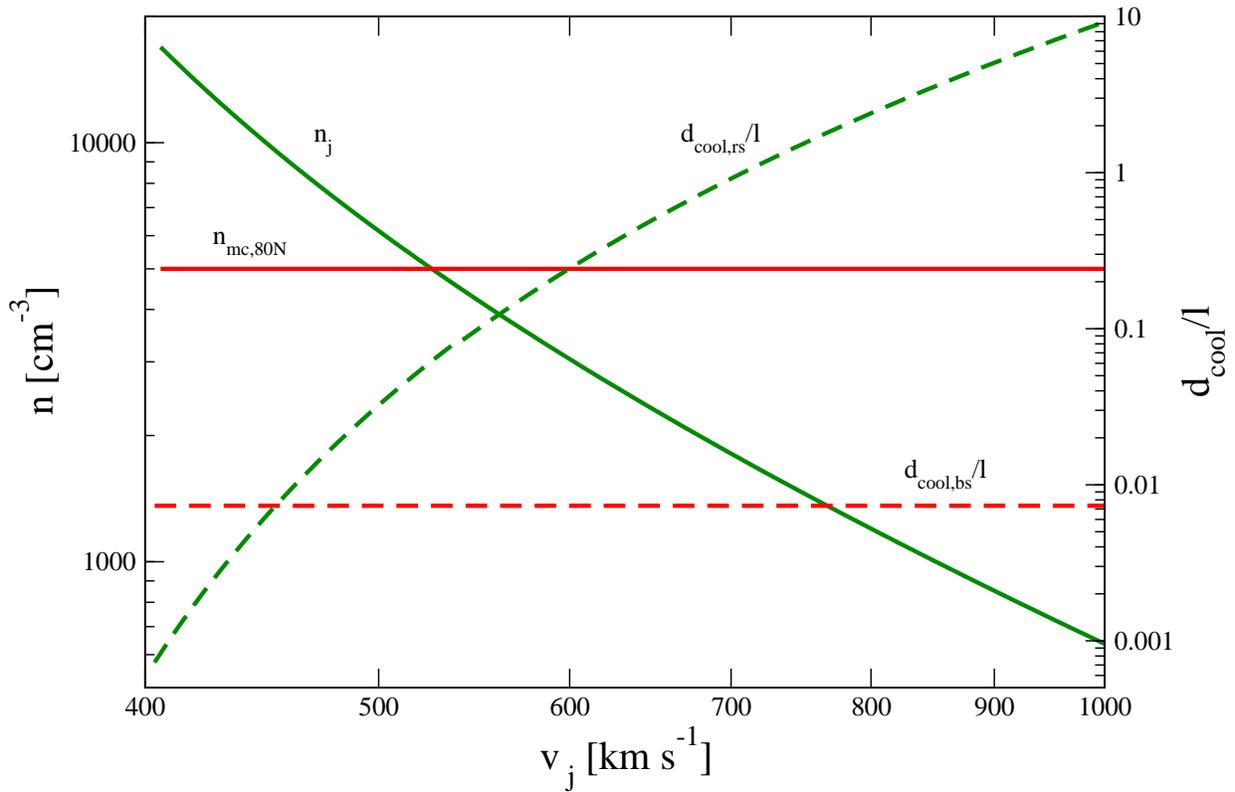}}
%\vspace{-0.7cm} 
\caption{\emph{Left axis:} Jet (green-solid line) and molecular cloud 
densities (red-solid line). \emph{Right axis:} Reverse- (green-dashed line) 
and bow-shock thermal cooling length (red-dashed line) in units of the source emitting size.  
The bow shock is radiative ($d_{\rm cool,bs}/l < 1$) for all possible values 
of $v_{\rm j}$ while the reverse shock is adiabatic ($d_{\rm cool,rs}/l > 1$) 
when $v_{\rm j} \gtrsim 800$~\kms.
\label{parameters}}
\end{center}
\end{figure}

The interaction of the jet with the ambient medium produces two shocks: a bow shock in the
molecular cloud, and a Mach disc (or reverse shock) in the jet. The shocked
matter from the molecular cloud and jet are separated by an (unstable)
contact discontinuity. This whole system propagates in the ambient
medium
at the bow shock speed $v_{\rm bs}$ \citep{blondin1989} that is assumed to be the 
derived velocities for the HH objects shown in Table~\ref{parameters}. Radiative shocks are those in which the plasma behind the shock emits thermal
radiation in a distance (i.e. also called the cooling distance, $d_{\rm cool} $) shorter to the jet radius, for which usually the source emitting size $l$ is adopted. In the opposite case,
$d_{\rm cool} > l$, the shock is non-radiative (i.e. adiabatic).

%In steady shocks,  the proper motions correspond to bulk motions of the 
%downstream plasma that produce the observed HH emission, instead of 
%morphological changes of this emission unresolved with our beam size. 
%This scenario applies for HH 80N because it is deeply embedded in 
%the cloud with constant confining pressure. However, the local 
%conditions of HH 80 and 81 are different than those of HH 80N.  

\citet{heathcote1998} pointed out that significant parts of HH 80 and 81 must be adiabatic. Also, these authors concluded that there are some caveats with the assumption of 
HH 80 and 81 as conventional bow shocks, specially the latter. Beyond HH 80, 
the jet escapes from the cloud and expands to a low density 
medium resulting in a large dispersion of the measured proper motions for
the fainter condensations (see Fig. 8 of \citealt{heathcote1998}). 
The derived velocity of HH 80 suggests that  
radio observations basically include emission from HH 80A, even though the emitting region of the HH object part escaping from the cloud could contribute in skewing our proper
motion determination to the south-west (i.e. the direction where the jet reaches the edge of the cloud). Probably, 
this effect tends to be more prominent with time. Similarly, in HH 80, there is a fainter condensation (HH 81B) located eastwards from the brightest one (HH 81A) and radio
observations possibly pick up the former. A possible explanation for the observed motion is a change in 
brightness of HH 81A relative to HH 81B during the time baseline inspected in this study. This would shift the measured centroid of the emission to the east resulting in appearance that the bulk motion is moving towards this direction. 
On the other hand,  
the local conditions of HH 80N, deeply embedded in 
the cloud with constant confining pressure, are different than those of 
HH 80 and 81. Below we discuss the different possible scenarios to explain the jet interaction with the cloud and the produced emission at HH 80N. In particular, we will consider that the synchrotron emission at 6~cm is produced by relativistic
electrons with energies $\sim 60 m_ec^2(B/{\rm mG})^{-0.5}$ in a magnetic field
$B$.

In Figure~\ref{parameters} we show the dependence between jet parameters estimated at the HH 80N position. Considering  the density of the molecular cloud, $n_\mathrm{mc,80N}\simeq 5\times10^3$~cm$^{-3}$ 
(Masqu\'e et al., in prep.), and its bow shock
velocity $v_{\rm 80N} = 263$~\kms, the ambient shocked material 
emits thermal emission over a distance  
%the thermal cooling length in the shock downstream region is 
$d_{\rm cool,bs}\sim 22$~AU behind the shock
\citep{raga2002}.
Given that $d_{\rm cool}$ in the bow shock is much smaller than the emitting size $l\sim 3000$~AU
(red-dashed line in Fig.~\ref{parameters}), the bow shock must be radiative. 
In such a case, the synchrotron  emission at 6~cm  could be produced by the 
compression of cosmic rays in the molecular
cloud, as in the case of supernova remnants in the radiative stage
\citep{chevalier1999}, or  by particles accelerated via second-order Fermi 
acceleration \citep{ostrowski1999}.

%\ll l$ 

 The HH~80N synchrotron emission can be also produced in the 
jet in the reverse shock region (Mach disc). 
%if it is strong enough to accelerate 
%particles via diffusive shock acceleration (Bell 1978). 
Considering the jet velocity $v_{\rm j}$
as a free parameter, we calculate the total jet density
$n_{\rm j}\sim n_{\rm mc,80N}/(v_{\rm j}/v_{\rm 80N}-1)^2$ (green-solid line in  
Fig.~\ref{parameters}) and  the reverse shock velocity for HH 80N as
$v_{\rm rs,80N}\sim v_{\rm j} - (3/4)v_{\rm 80N}$. Therefore, comparing the thermal
cooling length downstream the Mach disc ($d_{\rm cool,rs}$) with $l$, we
can see in Fig~\ref{parameters} (green-dashed line) that the Mach disc 
is non-radiative
($d_{\rm cool,rs} > l$) when $v_{\rm j}\gtrsim 800$~\kms. These values fall within the range of values found by \citet{marti1995} (600-1400~\kms) that are possibly representative of the jet velocity. As seen in the figure, this implies jet densities of $\lesssim 1000$  \cmt\ (but probably grater than 500 \cmt). The inferred large jet velocities and densities are the expected for a jet emanating from a high mass protostar. Thus, the scenario of having synchrotron emission in the reverse shock of the jet is the most likely.
In this case, 
relativistic electrons are accelerated  via 
the Fermi I acceleration mechanism  \citep{bell1978}
and emit synchrotron emission with an spectral index
$\le -0.5$, being the observed flat spectrum in HH~80N ($\alpha \sim -0.3$)
the result of thermal contamination. Assuming that relativistic electrons follow a power-law energy 
distribution $\propto E^{-2}$ that terminates at
$E_{\rm min} = 10m_ec^2$, the magnetic field in 
equipartition with non-thermal electrons and protons is $B_{\rm eq}\sim 1$~mG. If we adopt the energy density of non-thermal particles $U_{\rm nt}$ as  
$B^2/(8\pi)$, we found that only a small fraction ($\sim1$\%) of the jet kinetic energy density $U_{\rm kin} \sim 5\times10^{-6}(n_{\rm j}/500\,{\rm cm^{-3}})(v_{\rm j}/1000$~\kms$)^2$~erg~cm$^{-3}$, is converted into non-thermal energy in the shock downstream region \cite[e.g.][]{park2015}.

\section{Summary}

We carried out 6 cm observations of the HH 80/81/80N radio-jet  with the JVLA at C configuration. These observations, combined with previous VLA observations obtained with the same observing band and configuration than ours, allow us to derive proper motions for the HH 80, HH 81 and HH 80N radio-knots of the jet. The measured proper motion of HH 80N suggests that it belongs to the HH 80/81/80N radio-jet constituting the northern counterpart of the HH 80 and 81 objects. We also confirm the presence of Source 34 previously reported by  
\citet{masque2012b}. This supports the possibility that the jet extends further north of HH 80N. The derived velocities for HH 80 (351~\kms), HH 81 (223~\kms) and HH 80N (263~\kms) are quantitatively slower than those derived by \citet{marti1995} for the inner knots of the jet (600-1400~\kms) corresponding to very recent ejections. This implies that the cloud medium contributes significantly in slowing down the jet material at the position of these HH objects. Indeed, the measured direction of the proper motions of HH 80 and HH 81 makes evident the importance of the medium in the emission of these HH objects: they are located in the edge of the cloud and possibly expanding to a low pressure medium. As a consequence, they suffer morphological changes that could alter our proper motion determination. On the other hand, HH 80N is deeply embedded in the cloud with a more constant confining pressure.  Estimating the nature of the shock of HH 80N we found that it must be adiabatic and its emission is partly synchrotron radiation produced by electrons accelerated up to relativistic velocities in the reverse shock in the jet. We argue that the jet velocities must be $\gtrsim 800$~\kms, the derived jet density is  $\lesssim1000$~cm$^{-3}$ and the magnetic field in the shocks is $\sim1$~mG. The work presented here provides further evidence for particle acceleration in HH 80, 81 and 80N that produces the observed non-thermal emission.

\begin{table}[ht]
\footnotesize
\begin{center}
\caption{Parameters at 6~cm of selected radio knots.\label{propermotions}}
\begin{tabular}{p{1.5cm}ccccccc}
\hline
\hline
&   \multicolumn{2}{c}{Peak Position$^{\mathrm{a}}$}  & $S_\mathrm{\nu}$(6~cm)$^{\mathrm{a}}$ & \multicolumn{2}{c}{Displacement$^{\mathrm{b}}$}   &  Velocity $^{\mathrm{c}}$ & P.A.$^{\mathrm{d}}$ \\
Source   & $\alpha$~(J2000) & $\delta$~(J2000) & (mJy) &   $\Delta_{\alpha}('')$ &  $\Delta_{\delta}('')$ & (km~s$^{-1}$)   &   (deg.)      \\
\hline
HH 80 &  $18^\mathrm{h}19^\mathrm{m}06^\mathrm{s}04$ & $-20^\circ51'51
\rlap{$''$}.09$ & $1.3 \pm 0.1$& $ -0.75 \pm 0.30 $ & $ -0.74 \pm 0.32 $ & $ 351 \pm 104 $ & $ -135  \pm 28 $\\
HH 81 & $18^\mathrm{h}19^\mathrm{m}06
^\mathrm{s}69$ & $-20^\circ51'06
\rlap{$''$}.11$ & $ 1.1 \pm 0.1$ & $ 0.65 \pm 0.25 $& $ -0.12 \pm 0.30 $ & $ 223 \pm 85 $ & $ 101  \pm 38 $\\
IRAS~18162  &$ 18^\mathrm{h}19^\mathrm{m}12
^\mathrm{s}08 $& $-20^\circ47'31
\rlap{$''$}.00$ & $ 4.9 \pm 0.1$ & $ 0.07 \pm 0.20 $ & $ 0.17 \pm 0.28 $ &
$ 63 \pm 89 $ & $ 157  \pm 91 $\\
HH 80N &  $ 18^\mathrm{h}19^\mathrm{m}19
^\mathrm{s}75$ &$ -20^\circ41'34
\rlap{$''$}.85 $& $2.3 \pm 0.1$ &$ 0.13 \pm 0.30 $ &
 $ 0.78 \pm 0.211 $ & $ 263 \pm 71 $ & $ 10  \pm 36 $\\
\hline
\end{tabular}
\vspace{0.5cm}\\
\end{center}
$^{\mathrm{a}}${ Derived from Gaussian fits to the sources of the 6~cm map obtained from the 2013 data convolved to $6.0\arcsec \times 4.0\arcsec$ (PA = -14$^\circ$) as the resulting beam. }\\
$^{\mathrm{b}}${Observed angular displacement on the sky plane derived by comparing the 1989.8 and 2013.8 data (i.e. separated by 24 years) in the direction of the $\alpha$ and $\delta$  axes. Errors include the uncertainties of the alignment process of the maps of the two epochs. We excluded Source 34 from the table as it was observed only in the 2013 epoch. 
}\\ 
$^{\mathrm{c}}${ For an adopted distance of 1.7 kpc. 
}\\
$^{\mathrm{d}}${Position angle, measured from north to east.}\\
\end{table}

\end{document}